\documentclass[manuscript,screen]{acmart-hack}

\usepackage{amsthm,amsmath}
\usepackage{hhline}    

\usepackage{mathtools}
\usepackage{adjustbox}
\usepackage{caption}
\usepackage{MnSymbol}
\usepackage{paralist}

\usepackage{mathpartir}

\usepackage{mymacros}

\usepackage{xparse}
\usepackage{instructions}

\usepackage{wrapfig}	
\usepackage[caption=false]{subfig}

\usepackage{lang}

\usepackage{booktabs}   

\usepackage{fp}

\newlength\BARSIZE  
\newcommand{\inlinechart}[2]{%
\FPeval{\BLACKBARSIZE}{#1/#2}\textcolor{black!80}{\rule{\BLACKBARSIZE\BARSIZE}{1.6ex}}%
\FPeval{\BLACKBARSIZE}{1 - (#1/#2)}\textcolor{black!10}{\rule{\BLACKBARSIZE\BARSIZE}{1.6ex}}%
}

\newcommand*{\smallbar}[2]{%
$#1/#2$\hspace*{0.5ex}%
\the\numexpr#1*100/#2\%\hspace*{0.5ex}%
\inlinechart{#1}{#2}%
}

\newcommand*{\smallbarB}[3]{%
$#1/#2$\hspace*{0.5ex}%
\the\numexpr#1*100/#2\%\hspace*{0.5ex}%
\inlinechart{#3}{#2}%
}

\makeatother

\setcopyright{none}
\copyrightyear{2023}           

\usepackage{lineno}
\usepackage{url}

\begin{document}

\title[Program Analysis for High-Value Smart Contract Vulnerabilities]{Program Analysis for High-Value Smart Contract Vulnerabilities: Techniques and Insights}

\author[Y.Smaragdakis]{Yannis Smaragdakis}
\affiliation{
  \institution{Dedaub and University of Athens}
  \country{}
}
\email{yannis@dedaub.com}          

\author[N.Grech]{Neville Grech}
\affiliation{
  \institution{Dedaub and University of Malta}
  \country{}
}
\email{me@nevillegrech.com}          

\author[S.Lagouvardos]{Sifis Lagouvardos}
\affiliation{
  \institution{Dedaub and University of Athens}
  \country{}
}
\email{sifis.lag@di.uoa.gr}          
\author[K.Triantafyllou]{Konstantinos Triantafyllou}
\affiliation{
  \institution{Dedaub and University of Athens}
  \country{}
}
\email{kotriant@di.uoa.gr}            
\author[I.Tsatiris]{Ilias Tsatiris}
\affiliation{
  \institution{Dedaub}
  \country{}
}
\email{i.tsatiris@di.uoa.gr}          
\author[Y.Bollanos]{Yannis Bollanos}
\affiliation{
  \institution{Dedaub}
  \country{}
}
\email{ybollanos@dedaub.com}          
\author[T.Valentine]{Tony Rocco Valentine}
\affiliation{
  \institution{Dedaub and University of Malta}
}
\email{tvalentine@dedaub.com}          

\authorsaddresses{}

\begin{abstract}
  A widespread belief in the blockchain security community is that
  automated techniques are only good for detecting shallow bugs,
  typically of small value. In this paper, we present the techniques
  and insights that have led us to repeatable success in automatically
  discovering high-value smart contract vulnerabilities.  Our
  vulnerability disclosures have yielded 10 bug bounties, for a total of over \$3M,
  over high-profile deployed code, as well as
  hundreds of bugs detected in pre-deployment or under-audit code.

  We argue that the elements of this surprising success are a) a
  very high-completeness static analysis approach that manages to
  maintain acceptable precision; b) domain knowledge, provided by
  experts or captured via statistical inference. We
  present novel techniques for automatically inferring domain
  knowledge from statistical analysis of a large corpus of
  deployed contracts, as well as discuss insights on the
  ideal precision and warning rate of a promising vulnerability
  detector. In contrast to academic literature in program analysis, which
  routinely expects false-positive rates below 50\% for publishable results,
  we posit that a useful analysis for high-value real-world vulnerabilities
  will likely flag very few programs (under 1\%) and will do so with a high
  false-positive rate (e.g., 95\%, meaning that only one-of-twenty human
  inspections will yield an exploitable vulnerability).


\end{abstract}





\maketitle






\section{Introduction}

Smart contracts---autonomous programs deployed and operating on the
blockchain---present a particularly interesting setting for software
verification and validation approaches. The reason is stringent
correctness requirements, largely arising due to the large and direct
impact of possible contract errors. Smart contracts manage monetary
assets, often in the many millions or billions of dollars. The most
common domain for smart contracts deployed in the past few years has
been Decentralized Finance (DeFi), i.e., autonomous protocols that
implement many of the functionalities of conventional finance (e.g.,
lending, exchanging, options trading), often in innovative ways.

With this much real-world value, one might expect that cutting-edge
research techniques on program analysis and testing would have a
significant practical impact on smart contracts. This has arguably not
been the case, however.  \citet{263870} conduct a thorough study of
tens of thousands of Ethereum contracts reported vulnerable by six
leading academic projects. They find that under 2\% of the contracts
and only 0.27\% of the funds held in them have actually been
exploited. A very small part of this impact is explainable by the
tools having higher-than-reported false-positive (i.e.,
\emph{imprecision}) rates.  Instead, the main reason for this striking
result is a strong bias in the sample: contracts that hold funds are
very heavily scrutinized and much more likely to be false positives in
the analysis. Even analyses with 90\% precision (i.e., true-positive)
rates, in the overall contract population, have extremely high
false-positive rates in the subset of contracts that truly currently
manage funds.

Therefore, it should come as no surprise that program analysis tools
are not considered very valuable for smart contract security analysis.
A recent quote (one of many) by a prominent security analyst in the
Ethereum space captures the prevailing view: ``\emph{for an
experienced contract author, it's never the automated tooling that
finds the bugs that kill them}''~\cite{ethsecurityQuote}. Having a
practical impact on deployed contract security via automated techniques
seems like a formidable challenge.

In this paper, we present the techniques and insights that we have
used to meet this challenge. Although human understanding is essential
for finding many of the high-value vulnerabilities in deployed smart
contracts, our work shows that automated analyses \emph{can} pinpoint
high-value vulnerabilities.  The statement is more than a near-trivial
existence proof: it is based on \emph{repeatable} results, that have
yielded hundreds of vulnerability reports, as well as ten independent
bug bounties, for a total of over \$3M, over DeFi protocols managing
billions in assets.

The technical core of our approach consists of two main parts. The
first is \emph{symbolic value-flow (symvalic)} analysis: a
high-precision/high-completeness \emph{static analysis} approach,
whose technical elements are described in detail in a recent
publication~\cite{10.1145/3485540}. The second part is a \emph{corpus
analysis}: a complementary (to the static analysis)
\emph{statistical analysis} of irregular or unusual code patterns,
relative to a large number of other deployed smart contracts.

An interesting element concerns the interplay of symvalic (i.e.,
static) analysis and corpus (i.e., statistical) analysis. The
vocabulary of the statistical analysis is directly informed by
properties established by the static analysis. For instance, a common
concept over which anomalous code patterns can be detected is that of
\emph{tainting}: e.g., the combined analysis may ask
\begin{quote}
  ``can an untrusted user \emph{taint} (i.e., control) the $n$-th
  argument of an external DeFi protocol call \emph{in this contract},
  when that argument is typically determined to be \emph{untainted} 
  \emph{in many other contracts} that call the same DeFi protocol?''
\end{quote}

Finally, we present insights regarding the expected performance
metrics of a program analysis for detecting high-value
vulnerabilities.  We argue that a useful analysis for complex
vulnerabilities is unlikely to be very precise in terms of
conventional metrics---and that is fine! A false-positive rate of 98\%
(1 of 50 warnings being truly exploitable) may be prohibitive for a
top-tier academic publication, but it is perfectly expected when the
domain of evaluation is not controlled benchmarks but actual,
deployed, and scrutinized smart contracts. A warning rate of 0.5\%
(flagging one-of-every-200 contracts for a given vulnerability) may
seem entirely unimpactful in terms of a research evaluation of the
analysis benefit. (``99.5\% of the programs are already correct,
what is the need for an analysis to get this number to 100\%?'') But
reality is quite the opposite: a 0.5\% warning rate is entirely in the
credible range for high-impact vulnerabilities.

\section{Background}

\paragraph{Static Analysis.}
Static analysis is distinguished among program analysis techniques
(e.g., \emph{testing}~\cite{MeyerTesting,mci/Czech2016}, 
\emph{model checking}~\cite{Jhala:2009, Clarke:1986}, or \emph{symbolic execution}~\cite{King:1976, symbexsurvey})
by its emphasis on \emph{completeness}, i.e., its attempt to model all
(or, practically, as many as possible) program behaviors.  To achieve
this goal under a realistic time budget, static analysis often has to
sacrifice some \emph{precision}: the analysis may consider
combinations of values that may never appear in a real execution.
Maintaining high precision, while achieving completeness and
scalability, is a fundamental conceptual challenge. Algorithmic design of
static analyses has been trying to address this challenge for several decades,
with thousands of different research innovations.

All static analysis design is, therefore, an effort to simultaneously
achieve as much practical \emph{precision} and \emph{completeness} as
possible. (The third dimension of analysis quality, \emph{scalability}
can mostly be ignored for the purposes of our presentation: we restrict our
attention to algorithms that scale well, i.e., can give results in
realistic time frames---e.g., in the hours---under current technology.)
Generic definitions for the two terms can only be approximate, but,
intuitively:
\begin{bullets}
  \item an analysis is precise to the extent that the program behaviors
    it predicts are actually possible in program execution;
  \item an analysis is complete to the extent that the actually
    possible program executions are predicted.
\end{bullets}

Perfect precision and completeness is provably impossible, by standard
undecidability results over program reasoning. Therefore a static
analysis is by definition a game of approximation. Precision and
completeness are typically measured experimentally (or one of them is
guaranteed by-design, with the other measured experimentally). It
should come as no surprise that an analysis picks what to approximate
with more or less fidelity. For instance, the best known family of
static analyses, \emph{data-flow analyses} in conventional compilers,
will commonly treat all code conditions as both satisfiable and negatable,
without more complex reasoning.

Consider the example 
function below:\footnote{We use the Solidity language for code
examples. Solidity is dominant, accounting for more than 99\% of
deployed smart contracts on Ethereum, the largest programmable
blockchain.}
\vspace{1mm}\begin{javanonumbercode}
function whichPaths(uint x) public (returns uint y) {
  y = 3;
  if (x 
    y++;
  }
  if (x 
    y = y * y;
  }
}
\end{javanonumbercode}

In this function, the actual possible values for return variable
\sv{y} are \sv{3}, \sv{16}, and \sv{9}: it is not possible to satisfy
the first conditional but not the second. A static analysis, however,
may legitimately choose to consider all paths (i.e., all combinations
of branches in the program control flow) to be realizable, sacrificing a
bit of precision. Such an analysis could well predict that \sv{4}
(i.e., satisfying the first conditional but not the second) is a
possible return value. This approximation loses precision in this
example but gains the benefit of completeness when faced with other
complex conditions. (Additionally, the approximation most likely
is much more scalable than approaches that attempt full reasoning
over symbolic conditions.)

    
\paragraph{Analysis of Smart Contracts.}    
The domain of Ethereum smart contracts represents a high practical but
also intellectual challenge for program analysis. The analysis should
be extremely precise, so that the warnings for high-value contracts
are likely true positives that humans may have missed, and yet fairly
complete, since catching the easy ``certain'' cases is likely to yield
no warnings for contracts that truly manage funds. Thankfully, there
are elements of the domain that help.  First and foremost, smart
contracts are isolated from each other and coded defensively. This
introduces a high degree of modularity: the contract can be analyzed
mostly in isolation from others.  Nearly anything that comes from the
outside world is untrusted, unless either the data or the sender are
vetted through specific mechanisms. Second, the contracts are of
modest size: a deployed smart contract is at most of 24KB in binary
size. This corresponds to at most a few thousand lines of code. Common
sizes of ``large'' smart contracts are under 1KLoC, with another 1KLoC
inherited or called in libraries.

The small size and (relative) modularity of smart contracts means that
we can apply analysis techniques that are more ambitious (in terms of
precision) than in a general-purpose language setting. Past work has
used ambitious program reasoning
techniques~\cite{9152791,grossman18,grossman20} (indeed, even full
program verification using proof assistants and off-line logics has
been employed~\cite{certora,8429306}). Our work is explicitly about
\emph{automated} techniques that apply universally, without needing
human input (i.e., specifications) or per-smart-contract
customization.

Our approach aims to be highly complete without sacrificing (much)
precision. Completeness problems are often brought up in the domain of
smart contracts. For instance, the Manticore symbolic execution
framework~\cite{manticore} (one of the foremost for Ethereum smart
contracts) ``achieves on average 66\% code
coverage''~\cite{Manticore66}.  This prompted the Trail of Bits
security firm's company account to tweet ``\emph{Why can't a symbolic
executor achieve 100\% coverage in a teensy little smart
contract?}''\cite{ToBTweet}, capturing the frustration of consumers of
analysis results.

\section{Symbolic-Value-Flow Static Analysis}

Symbolic-Value-Flow (``symvalic'') analysis is the first weapon in our arsenal.
We next present an informal overview, highlighting
its design principles.

\subsection{Overview}

Symbolic-Value-Flow static analysis
is much like a common inter-procedural data-flow analysis: a
``value-flow'' analysis, such as a points-to or a taint
analysis. Being a value-flow analysis implies that every variable is
statically assigned a finite set of values and a fixpoint computation
grows the finite sets according to monotonic equations.

In symvalic analysis, the values can be both concrete (e.g., numbers)
and entire symbolic expressions. For instance, consider the example function below:

\vspace{-4mm}\begin{javanonumbercode}
  function sensitive(address recipient) public {
    require(authorized[msg.sender]);
    selfdestruct(recipient);
  }      
\end{javanonumbercode}

The analysis will consider a set of concrete and symbolic values for
all external input variables. For instance, for numeric variables, the
analysis considers small constants (0,1, and up to 3 constants under
256), large constants (to cause overflow), and a pseudo-random choice
of constants from the program text.

For external inputs of
``contract address'' type, such as the \sv{msg.sender}
implicit argument, the analysis will consider values that include:
\begin{itemize}[$\bullet$ ]
\item constants in the contract text that resemble addresses (i.e., 160-bit integers)
\item the symbolic values \sv{<<owner>>} and \sv{<<unprivileged-user>>}.
\end{itemize}

Similarly, the values initially considered for the \sv{recipient} argument include:
\begin{itemize}[$\bullet$ ]
\item constants in the contract text that resemble addresses
\item the symbolic values \sv{<<owner-unique-value>>} and 
  \sv{<<user-unique-value>>}.
\end{itemize}

The difference between the symbolic values is that the former
(\sv{<<owner>>} and \sv{<<unprivileged-user>>}) will be considered
\emph{bound-variables} for symbolic reasoning purposes: although we
treat them symbolically, the caller cannot set them freely. In
contrast, the latter (\sv{<<owner-unique-value>>} and
\sv{<<user-unique-value>>}) are free variables and symbolic reasoning
can propose concrete values for them, in order to solve constraints.
Section~\ref{sec:dependencies} will describe in more detail how these
values are \emph{dependent}: the analysis will model separately the
case of the current caller of the contract being \sv{<<owner>>} and
that of it being \sv{<<unprivileged-user>>}, and similarly for the
values they supply to arguments.

Symbolic values propagate and are used to form more complex
expressions, whereas concrete values get constant-folded. For
instance, in the example, the program expression
\sv{authorized[msg.sender]} is a lookup in a mapping structure on
Ethereum storage (i.e., analogous to the heap in a conventional
language). Due to the way Ethereum storage is organized, the
expression is loading from the symbolic address \sv{SHA3(<<owner>> ++
  0x0)}---\sv{SHA3} is the Keccak-256 hash function, \sv{++} a byte
array concatenation operator, and \sv{0x0} (i.e., zero) the constant
offset that identifies the \sv{authorized} mapping among other
attributes in storage. This symbolic expression is propagated by the
analysis as a value for the corresponding local variable (unnamed in
the source language but present in the intermediate representation).

\begin{figure*}
  \centering
  \includegraphics[width=0.85\textwidth]{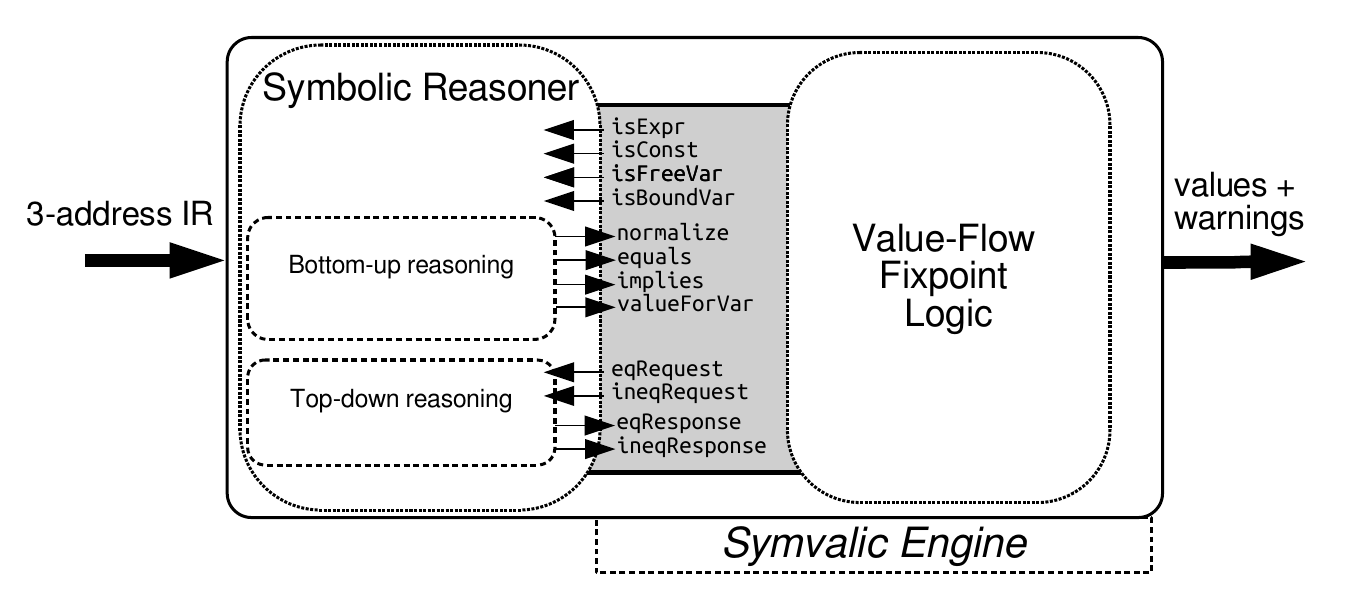}
  \caption{Symvalic Analysis Architecture: API (slightly simplified) 
    between symbolic reasoning and value propagation shown.}
\label{fig:architecture}
\end{figure*}

The analysis proceeds via a close interaction of a value-flow
fixpoint loop and a symbolic reasoner.  Figure~\ref{fig:architecture}
shows the overall architecture, including the symbolic reasoning
component and the value-flow component, as well as the interface
between them and the two sub-components of symbolic reasoning: a
bottom-up and a top-down component.
The full interface and bottom-up/top-down distinction are elements of
importance mostly for practical concerns, especially
efficiency. 

If viewed as an idealized reasoner over infinite expressions (i.e.,
without worrying about computational efficiency) the main products
of the symbolic reasoner are:
\begin{itemize}[$\bullet$ ]
\item a predicate \sv{normalize(expr, normExpr)} that returns for each
  expression its minimal equivalent form. This applies to both
  arithmetic and logical expressions;
\item a predicate \sv{implies(exprStrong, exprWeak)} that checks whether
  one logical expression implies another;
\item a predicate \sv{valueForVar(var, expr)} that proposes values
  (i.e., symbolic expressions) for free variable \sv{var}, so that
  logical conditions get satisfied. The solver proposes such solutions
  in bulk, for any interesting logical expressions (the main case
  being equalities, which are usually much harder to satisfy than
  inequalities) and the analysis filters the solutions it chooses to
  truly try---e.g., if they satisfy an equality that was previously
  not satisfied. 
\end{itemize}

\noindent Inevitably, all three predicates are incomplete, i.e., will not
capture all (infinite) true relationships.

The analysis appeals to these predicates throughout normal value
propagation. Symbolic values are continuously \emph{normalized} (i.e.,
simplified up to minimal form) and are also used to satisfy
control-flow constraints, i.e., predicates in conditional statements.
This makes the analysis path-sensitive: the information pertaining to
a program statement is derived to be compatible with the conditions
necessary for reaching the statement. Mechanism-wise, this is a part
of the \emph{dependencies} machinery, described in
Section~\ref{sec:dependencies}.

The close interaction of value-flow and symbolic reasoning is a rather
unconventional aspect of symvalic analysis. This is in contrast to
explicit invocation of solvers in a static
\nocite{christakis}
analysis~\cite{ethor,maian,teether}. It is common for a static analysis to
collect a large condition and dispatch an external solver (e.g.,
Z3~\cite{z3}) to satisfy the condition. Symvalic analysis, however,
invokes the symbolic solver a lot more regularly, to continuously
simplify and solve conditions. Such appeals to the symbolic solver are
essential, in the heart of the analysis.

\subsection{Precision and Dependencies}
\label{sec:dependencies}

\newcommand{\dependencies}[1]{\textbf{$\langle$}#1\textbf{$\rangle$}}
\newcommand{\VarMayBe}[3]{\sv{#1} \rightarrow #2\  \dependencies{#3}}
\newcommand{\VarMayBeNoDeps}[2]{\sv{#1} \rightarrow #2}
\newcommand{\combine}{ \oplus }

Symvalic analysis is a static analysis, aiming for completeness, i.e.,
covering \emph{many} program behaviors. However the analysis is
neither sound nor complete: it can both model unrealizable behaviors
and miss actual behaviors. The analysis intends to strike a good
balance between \emph{completeness} and \emph{precision}: to cover
most program behaviors, while still predicting
behaviors that are realizable to a large degree.

To achieve precision, while keeping a set-based treatment, symvalic
analysis associates variable-value tuples (as well as statements in
the code) with \emph{dependencies}.  Every analysis inference of the
form ``variable \sv{v} may have value (i.e., concrete or symbolic
expression) $e$'' (as well as of the form ``this statement is
reachable'') is associated with \emph{sets of mappings of variables to
  values} that have led the analysis to make the inference.

We distinguish two kinds of dependencies: \emph{local} (reset
per-function) and \emph{transaction} (i.e., current-execution)
dependencies. We write $\VarMayBe{v}{e}{{d}^L;{d}^T}$ to designate that
variable \sv{v} may hold value $e$ under local dependencies ${d}^L$
and transaction dependencies ${d}^T$. (When there is no need to distinguish
local and transaction dependencies, we write $\VarMayBe{v}{e}{D}$.)
To illustrate dependencies,
consider a simple contract fragment:

\vspace{-3mm}\begin{javanonumbercode}
contract Safe {
 address owner;   // set at construction
 mapping (address => uint) public balanceOf;
 function deposit(address to,uint amount) public {
   require(msg.sender == owner);
   uint curBalance = balanceOf[to];
   uint nextBalance = curBalance + amount*90/100;
   balanceOf[to] = nextBalance;
 }
}
\end{javanonumbercode}

Let us focus on the value of variable \sv{nextBalance} during an
invocation \sv{deposit(0x42, 200)}.\footnote{Contract addresses are
  160-bit integers and generated randomly. We show a very unlikely
  short address (\texttt{0x42}) for the sake of conciseness.}  For \sv{nextBalance} to even have a
value, its assignment statement needs to be reachable. Therefore the
``\sv{require(msg.sender == owner)}'' statement needs to be
satisfied. This forces the mapping $[ \sv{sender} \rightarrow
  \sv{<<owner>>} ]$ to appear in the transaction dependencies. (The
symbolic variable \sv{sender} stands for the Solidity \sv{msg.sender}
expression and \sv{<<owner>>} is a symbolic value generated to
represent the contract that originally called the current contract's
constructor and, thus, set storage fields to initial values.)

Additionally, values of local variables will be parts of
the dependencies. Assume that at some point the analysis has inferred
that storage mapping \sv{balanceOf[0x42]} has two possible values: $1$
and $80$, and then considers the call \sv{deposit(0x42, 200)}.

The inferences for variable \sv{nextBalance} will then be:

\noindent $\VarMayBeNoDeps{nextBalance}{181}$
\textbf{$\langle$}
 \{[\sv{to} $\rightarrow$ \sv{0x42}],
      [\sv{amount} $\rightarrow 200$],
      [\sv{curBalance} $\rightarrow 1$] \} ; \{[\sv{sender} $\rightarrow$ \sv{<<owner>>}] \}
 \textbf{$\rangle$}\}

        and
        
\noindent $\VarMayBeNoDeps{nextBalance}{260}{}$
\textbf{$\langle$}
 \{[\sv{to} $\rightarrow$ \sv{0x42}],
      [\sv{amount} $\rightarrow 200$],
      [\sv{curBalance} $\rightarrow 80$] \}; \{[\sv{sender} $\rightarrow$ \sv{<<owner>>}] \}
  \textbf{$\rangle$}\}.

The first of the two inferences can be read as ``\sv{nextBalance} is expected to hold
the value $181$ in executions where \sv{to} holds value \sv{0x42}, \sv{amount}
holds value $200$, the most recent storage load into variable \sv{curBalance}
returned value $1$, and the transaction initiator (i.e., external caller) is \sv{<<owner>>}''.
The first three mappings make up the local dependencies for the inference, whereas
the last is the transaction dependency.

Dependencies are \emph{combined}, with an operator denoted
$\combine$. Combination of dependencies happens for every control-flow
or data-flow join point: when two branches merge, or when two values
are used as operands in the same operation. Combining two sets of dependencies is a check for
compatibility, to prevent mixing information from guaranteed-separate
executions. Combining dependencies succeeds if there are no
conflicting mappings for the same variable. For instance, the
two inferences above cannot be combined: they conflict on the mapping
for variable \sv{curBalance}. If dependencies do not conflict (thus,
combining them succeeds), combination is a mere pairwise union of the
mappings sets.

\subsection{Precision and Completeness}
\label{sec:soundness}

Using the \sv{Safe} contract example, we can illustrate why symvalic
analysis chooses to be neither sound nor complete, i.e., why it
accepts some incompleteness in order to be more precise, and why it
accepts less-than-full precision (e.g., false positives) in order to
be more complete or more scalable.

\paragraph*{Incompleteness} The last statement of function \sv{deposit} stores
\sv{nextBalance} back into the persistent storage mapping
\sv{balanceOf[0x42]}.  Therefore, new inferences are possible for
variable \sv{nextBalance}, which again cause new values to flow into
\sv{balanceOf[0x42]}, \emph{ad infinitum}. Static analysis typically
resolves this potentially infinite computation by overapproximation
(e.g., a finite-height lattice that joins values into abstract values,
at the expense of some loss of precision). In contrast, symbolic
execution resolves non-termination (arising in the case of looping) by
arbitrarily truncating execution traces (e.g., unrolling loops only a
small number of times, or executing a fixed number of total
instructions).

Symvalic analysis is agnostic regarding the handling of cyclic flow:
both overapproximation and finite truncation are acceptable, per case.
For instance, the current symvalic setting uses overapproximation for values
from some sources (e.g., environment variables, such as gas remaining,
current block number, miner of block). In most cases, however,
the analysis favors concrete numeric values (such as those in \sv{balanceOf[0x42]}),
which yield other concrete values, up to a \emph{finite} number of arithmetic
operations. This treatment is one that explicitly favors
precision over completeness: the analysis is not guaranteed to
model all values arising in real executions. The advantage, however,
is that the values that the analysis considers are \emph{likely} to be
realizable, i.e., precision is enhanced.

\paragraph*{Precision} Dependencies can lend arbitrary precision to
an analysis. Even fully concrete execution can, for instance, be
simulated by maintaining in dependencies all dynamic variables of an
execution. Conceptually, dependencies can be viewed as a generalization of
\emph{context-sensitivity}~\cite{Sharir:Interprocedural} in static analysis, or an
instance of relational analysis (e.g., \cite[Sec.4.4.1]{dvanhorn:Neilson:1999}, \cite[Sec.7]{relational}).
Such mechanisms group together dynamic executions for uniform static
treatment. Precision arises because executions mapped to different
groups are never confused. The challenge, however, is to maintain
sufficient precision without suffering from extreme lack of
scalability (termed the \emph{state explosion} problem). The state
explosion problem is the bane of fully-precise program analysis
approaches, such as concrete testing or concrete-state model checking.
The number of possible value combinations rises exponentially, per set
combinatorics. In the setting of symvalic analysis, if the
dependencies mechanism were to keep full concrete state (i.e., what
\emph{precisely} are the contents of storage or variables in a
simulated execution), the analysis would suffer tremendously in
scalability. In practice, even dependencies on a handful of variables
can render the analysis unscalable.

Symvalic analysis maintains a balance between precision and
performance by computing dependencies only on a small subset of
program variables. As a consequence, the analysis can produce warnings or values
that are \emph{imprecise}, i.e., do not correspond to actual executions.

These precision limits of dependencies in the analysis are as follows
(with current defaults listed in parentheses, for concreteness):

\begin{itemize}[$\bullet$ ]
\item A bounded number of arguments of the current function, such as
  \sv{to} and \sv{amount} in the example. (Currently: up to 3 arguments
  are kept as local dependencies.)
\item A bounded number of local variables that load values from shared
  memory (storage), such as \sv{curBalance} in the
  example. (Currently: the first variable loading from storage per
  function is kept in local dependencies.)
\item A bounded number of external arguments (i.e., arguments supplied
  at the original entry point of the transaction, by an external
  caller). (Currently: the first 2 arguments of the transaction entry
  point are kept in transaction dependencies. This case is not shown in
  the example. However, if we were to change the code to make function
  \sv{deposit}---which is a transaction entry point---call a
  different, internal function, the values of \sv{deposit}'s arguments,
  \sv{to} and \sv{amount}, would be kept in the transaction dependencies when
  analyzing the internal function.)
\item The transaction's current caller (\sv{msg.sender} in Solidity).
  (Currently: kept in transaction dependencies.)
\end{itemize}

\newcommand{\extPredicate}[1]{\textbf{\textsc{#1}}}

\newcommand{\StorageLocMayBe}[2]{\llbracket #1 \rrbracket \Rightarrow #2}
\newcommand{\StmtDeps}[2]{|#1|\  \dependencies{#2}}
\newcommand{\norm}[1]{\extPredicate{normalize}($#1$)}
\newcommand{\ruleAnd}{\hspace{0.5cm}}

\section{Corpus Analysis}

The symvalic analysis technology has been a great boon to our efforts
to have repeatable high impact on smart contract security. However,
the analysis by itself is just a way to get answers. An essential
element of the approach is to know to ask the right questions. For
instance, a very simple question may take the form ``\emph{is the first
argument of a \sv{swap} call tainted, i.e., controllable by an
untrusted caller?}'' That is, at the very least the analysis will need
to know which arguments of external calls are sensitive and can lead
to vulnerabilities. Other interesting questions concern missing
information (e.g., missing a network identifier or a nonce when
cryptographically signing messages), vulnerable code patterns (e.g.,
effects after a possibly-reentrant external call), and much more.

Formulating the right questions, i.e., writing symvalic analysis
\emph{clients}, requires significant domain expertise. Security experts,
with deep knowledge of protocols and programming patterns in the DeFi space
need to be involved. This effort needs to be ongoing, with new libraries and
protocols (raising new requirements for integrating with them) being
continuously introduced in the ecosystem.

Abstracting away for one step, the need for domain/security expertise
is really a modularity enhancement. We need domain expertise so that
the analysis can break up its effort in convenient single-contract
chunks while having a summary of what \emph{other} contracts expect
in terms of their interactions with the current contract-under-analysis.

The second technique that we employ intends to address exactly this
need.  \emph{Corpus analysis} is a summarization of the behavior of
contracts and a subsequent statistical analysis. This enables
extracting domain expertise without the help of human experts. For
instance, it is not necessary for an expert to indicate that the first
argument of a \sv{swap} call is sensitive: if analysis of hundreds of
past contracts shows that this argument is almost always
\emph{untainted} (i.e., only checked callers can supply it), the
corpus analysis will establish this property as a question for the
symvalic analysis of a contract, to report violations.

More specifically, corpus analysis is a summarization of contract
behavior in two ways:
\begin{bullets}
\item For a contract itself, several key behavioral properties
  are summarized, so that the contract's callers can be checked
  in conjunction with these properties.
\item For all the callers of a contract (more accurately, a
  specific API function), their ``usual'' behavior is summarized
  so that other callers can be checked for deviation from the
  common patterns.
\end{bullets}

Corpus analysis works in synergy with the static analysis of
a contract: it both consumes the results of static analysis and
produces results for the static analysis to consume.

Specifically, for each contract statically analyzed (via symvalic
analysis), a wealth of behavioral summaries get produced. Some of the
most important ones are listed below:
\begin{bullets}
\item Which functions can reach a \sv{delegatecall} instruction
  (which is a common way to yield full control and, thus, a common vector
  for vulnerabilities).
\item Which functions have which of their arguments control
  quantities with monetary significance (e.g., flowing to a money transfer
  source/destination/amount).
\item Which functions perform state initialization.
\item Which functions return values that can be manipulated by an
  external caller (i.e., values depending on storage state that any
  caller can affect---e.g., by depositing funds).
\item Which functions allow reentrancy, i.e., yield control to a
  party supplied as a parameter. This does not indicate that the
  function itself suffers from a vulnerability: the function may
  not have state changes after the yielding of control, or the function
  may be checking that it gets called only by a trusted caller. However,
  \emph{callers} of such a function may be vulnerable to reentrancy attacks.
\item Which functions perform checked money transfer operations, i.e.,
  examine the permissions of their caller. This implies that any
  caller of such a function, when propagating argument values to the function,
  should itself examine the permissions of \emph{its} caller.
\item Which functions perform guarded/unguarded (i.e., with checks as
  to the function's caller) external calls and to which other functions of
  external contracts.
\item Which functions make external calls of monetary significance and whether
  (and which of) the arguments of such calls are tainted/untainted.
\end{bullets}

The above summaries get statistically analyzed and \emph{re-imported}
for further inference. Note the recursive nature of many of the above
definitions. To produce results for tainted/untainted external calls
of monetary significance, the analysis needs to first receive
information as to \emph{which} external calls have monetary
significance. To produce functions that allow reentrancy, we need to
know whether the external functions \emph{they} call allow reentrancy.


The most useful statistical summarization results of corpus analysis
are some of the simplest: e.g., whether a specific argument of an API
is usually tainted/untainted; whether a given function is usually
called only by trusted/checked callers or by anyone.

Feeding back into the static analysis, the results of corpus analysis
help formulate the right analysis questions:

\begin{bullets}
\item Knowing that an external call allows reentrancy allows detecting
  \emph{transitive reentrancy} attacks. It is worth emphasizing again
  that the external function itself may not be vulnerable, even though
  it yields control to a third party: the function may be checking its
  caller, or its code may have no permanent effects after yielding
  control. However, any \emph{caller} of this function may be violating
  these properties, becoming transitively vulnerable.
\item Knowing that a specific API parameter is typically (based on
  statistics of common usage patterns) \emph{not} allowed to be
  controlled by untrusted callers can yield a high-precision warning
  if the current contract allows the same parameter to be tainted.
\item Knowing that an external call is statistically only reached after
  authorization checks over the caller can yield a high-precision warning
  if the current contract has no such checks.
\end{bullets}

Corpus analysis has, thus, helped us scale up vulnerability detection
by extracting domain expertise, instead of always relying on human experts.
Key to the approach is the availability of hundreds of thousands of smart
contracts already deployed on public blockchains. Secondarily the approach is
enabled by the composable nature of DeFi protocols, which means that many
different codebases use the same external protocols, so that common usage
patterns can be ascertained.

\section{Discussion and Insights}

Automated program analysis tools have an extremely rich history in
Computer Science, with entire research communities centered around
program analysis.  Security is often brought up as a major motivation
for program analysis research. And yet, in one of the most critical
security domains in existence, smart contract security, the impact of
automated analysis tooling is repeatedly estimated to be close to
zero. There are numerous possible explanations for this phenomenon,
such as: the speed of development in smart contract engineering, which
has not (up to now) allowed the time for research solutions to catch
up; the immense, disproportionate (relative to other software
engineering domains) investment on smart contract correctness that has
attracted top human expertise to smart contract auditing, leaving
little room for automated solutions; the fundamental limitations of
fully automated solutions to software correctness, relative to, e.g.,
automation-assisted program verification, under the
guidance/specification of human experts. (Although, on the latter
point, a lot more can perhaps be said about the fundamental
limitations of automation-assisted program verification.)

We would like to add to the discussion our observations on the
contrast of academic research in program analysis (an area in which we
have significant presence for over 15 years) and industrial impact.
The discord between the two can be well-captured by the distance
in answers to two questions.\footnote{The first author has repeatedly
asked these questions in seminar talks at top Universities and research
institutions, always with the same disparity between audience expectation
and reality.}

\begin{enumerate}
\item What \emph{warning rate} do you expect an impactful program analysis
  to have? I.e., out of 100 programs, how many do you expect to be flagged
  for vulnerabilities for an analysis to be useful?
\item What precision, in terms of \emph{false positive rate}, do you expect
  an impactful program analysis to have? I.e., out of 100 warnings, how many
  do you expect to indicate real vulnerabilities?
\end{enumerate}

Researchers in program analysis typically give answers in the range of \emph{10-30\%}
for the first question and \emph{below 30\%} for the second. Their expectation is
that an impactful analysis will flag many contracts and will do so with very high
precision.

Our own experience over random (de-duplicated) deployed smart
contracts in active use is that realistic answers for the first
question are \emph{in the 1\% or below} range and answers for the
second question are in the \emph{over-90\%} range. That is, the most
useful analyses flag very few contracts---around 1\% or fewer---and it
is fine if these warnings are often false positives. Anything outside
these ranges is suspiciously off: it is probably a low-value analysis
that is flagging too many contracts for ``vulnerabilities'' that human
experts consider innocuous.

The order-of-magnitude discrepancy between expectation and reality is
not that surprising, if one considers the setting. Program analysis
experts are used to having their work evaluated over injected
vulnerabilities, systematic benchmark suites with bug examples, or at
least less-scrutinized, error-prone code. In the reality of deployed
smart contracts, end-to-end vulnerabilities are extremely rare. A 95\%
false-positive rate might not be enough for a top-tier publication,
but in practice it is almost too good to be true! It means that a
vulnerability inspector will find a true, actionable vulnerability for
every 20 inspections.

Therefore, it is clear that we have two problems: one of wrong metrics
and one of perverse incentives (the latter being largely a consequence of the
former).

At first glance, it seems reasonable to consider an analysis with a
0.5\% warning rate to be less-than-impactful. After all, ``if 99.5\%
of programs are correct, what is the big deal about getting that number
to 100\%? The relative improvement is tiny.'' However, this exact
rationale leads to discounting the most practically impactful
techniques. Conversely, techniques that focus on shallow
``vulnerabilities'' score much higher than impactful techniques on
both metrics.

We believe that program analysis metrics should adjust to the rarity
of the phenomenon they seek to capture. More is not better, nor is
less.  Instead, closeness to reality or absolute end-to-end impact is
the only truly better measure for estimating the value of a
technique.

\section{Practical Impact}
Our analysis has been applied to all new contracts deployed on
the Ethereum blockchain since the beginning of 2021. The analysis has
flagged numerous exploitable vulnerabilities---e.g.,
\cite{primitive-postmortem,pawnstars,badbot,lookma,yieldskimming}.  We
have made several vulnerability disclosures, some of which resulted in
major rescue efforts~\cite{warroom,pawnstars}. The total funds under
threat from these vulnerabilities well exceed a billion dollars. We
have received ten independent bug bounties totalling over \$3M.


From a program analysis standpoint, many detected vulnerabilities have
the same general structure: they correspond to warnings of the form
``an \emph{untrusted caller} $C$ can reach argument \sv{X} of a
sensitive operation and supply parameter $Y$ that is tainted'' or
``an untrusted caller can reach a sensitive operation (at all)''. That
is, the vulnerability warnings typically query the main relations
produced by the analysis: $\VarMayBe{X}{Y}{* ; [\sv{sender} \rightarrow C]}$,
as well as $\StmtDeps{i}{* ; [\sv{sender} \rightarrow C]}$.
The untrusted caller $C$ corresponds
to symbolic value \sv{<<unprivileged-user>>}, as seen in earlier
examples, stored in the transactional dependencies of the symvalic
analysis. The tainted parameter value $Y$ is typically the symbolic
value \sv{<<user-unique-value>>}, mentioned earlier, which designates
that the value is completely unconstrained. The exact nature of the
sensitive operation with argument $X$ varies by vulnerability. For instance:

\begin{itemize}
\item \emph{transferFrom}~\cite{pawnstars}: the contract is authorized to manipulate
  the funds of some accounts, and its code allows a direct transfer of
  funds from a tainted source to a tainted sink.
\item \emph{swap}~\cite{lookma}: an exchange of funds from one token to another,
  (taking place after a loan and a liquidation of ``shares''). The
  taintedness of the token being swapped and of the amount
  swapped are essential to the attack.
\item \emph{flashswap}~\cite{primitive-postmortem}: code executed upon an external service's
  granting of a loan does not check that the loan parameters were
  as requested: the attacker can invoke the callback with tainted
  loan parameters (e.g., tainted token).
\item \emph{manipulated swap}~\cite{yieldskimming}: the contract periodically converts its
  profits, and anyone can invoke this functionality. The attack is
  based on the reachability of this code by untrusted callers, and not
  on taintedness. The attacker calls the functionality exchanging the
  funds after having manipulated the online prices of the exchange
  service doing the conversion.
\end{itemize}

As can be discerned from the above descriptions, the attack point is
buried deep in the code, under several complex conditions. One can,
therefore, see why the precision and
completeness of symvalic analysis is important for the detection of
the vulnerability. As we write in a vulnerability
report~\cite{pawnstars}: ``What made our symbolic-value flow analysis
useful was not that it captured this instance but that it
\emph{avoided} warning about others that were \emph{not}
vulnerable. The analysis gave us just 27 warnings about such
vulnerabilities out of the 40 thousand most-recently deployed
contracts!''

\section{Related Work}


\paragraph{Program Analysis for Ethereum Smart Contracts}

The small size and high value of Ethereum smart contracts has made them a suitable target
for applying a variety of program analysis techniques, resulting in a plethora of academic
\cite{oyente,Grech2018oopsla,Tsankov2018,maian,honeybadger,10.1145/3293882.3330560,
10.1145/3428258,ethainter,10.1145/3377811.3380334} and industrial \cite{slither,
10.1145/3395363.3404366,mythril,certora,manticore} tools emerging.
While most smart contract tools focus on vulnerability detection, past work has also
focused on empirically identifying optimization opportunities \cite{9026761},
gas cost estimation \cite{albert2020gasol} using recurrence-relation theories or
even superoptimization \cite{10.1007/978-3-030-53288-8_10} using SMT.

Tools \cite{oyente,maian,10.1145/3293882.3330560,manticore,teether,10.1007/978-3-030-41600-3_11} applying symbolic
execution techniques have been very popular due to the precision of their reported warnings.
(And also, a cynic might remark, their ease of implementation on a platform where
the source language changes constantly and the low-level IR is extremely hard to analyze.)

Additionally, approaches \cite{Grech2018oopsla,Tsankov2018,ethainter,slither,10.1145/3428258,
Albert2018,vandal-tr} based on static analysis have seen success due to their high completeness
and scalability. Even though conventional static analysis tools \cite{ethainter} have achieved high
precision by tuning the analysis to common programming patterns found in Ethereum smart contracts, symvalic analysis offers a more precise analysis
while being agnostic to these specific program patterns.

Fuzzing-based tools \cite{10.1145/3377811.3380334,christakis,10.1145/3368089.3417064,
10.1145/3319535.3363230,10.1145/3238147.3238177} have also been successful in precisely
reporting smart contract vulnerabilities. Notably, the recently presented Harvey fuzzer \cite{christakis}
combines static analysis with fuzzing by using static analysis to guide a greybox fuzzer.

\paragraph{General Program Analysis and Verification}

Relational analysis techniques have been successfully applied in recent years,
to tackle the problem of JavaScript static analysis, where precision is
critical to getting a useful analysis. \emph{Value partitioning}~\cite{valuepartitioning2020},
is an efficient trace partitioning~\cite{tracepartitioningabsdomain} variant,
where the analysis does not attempt to refine abstract states, but instead, abstract values.
This apporach manages to circumvent the expensive abstract state partitioning~\cite{weaklysens}
or additional backwards analysis~\cite{ddvaluerefinement} that previous apporaches required,
while maintaining precision.

Symbolic execution has seen numerous variations that offer a different
balance of scalability, completeness, and
precision. \emph{Compositional} symbolic
execution~\cite{godefroid07compositional,anand-compositional} has
attempted to address scalability issues by use of summaries.
\emph{Steering}
techniques~\cite{10.1145/2393596.2393636,10.1145/2509136.2509553}
attempt to achieve higher coverage or depth, e.g., by prioritizing
paths that are yet unexplored. Symvalic analysis has similar goals,
but its coverage is only a small part of the story: as a static
analysis, it explores many values at once and coverage is only
incidental. At the same time, it may suffer from higher imprecision
than symbolic execution techniques, since precision is limited by its
current dependencies scheme.

Accordingly, symvalic analysis can be viewed as an attempt to address
the state explosion problem.  The model checking literature contains
several approaches with similar goals, ranging from compositional
\emph{assume-guarantee reasoning}~\cite{10.1145/151646.151649} to
symmetry reduction~\cite{NORRISIP199397,symmetry-emerson}, to partial
order reduction~\cite{10.1145/1040305.1040315}. Symvalic analysis uses
a very different scheme, due to both symbolic reasoning and its
controlled sacrifice of precision. Conceptually, the combination of
abstract interpretation and model checking (e.g.,
\cite{10.1145/186025.186051}) has a similar flavor, but the actual
mechanisms differ substantially.

Whitebox
\cite{SAGECACM,godefroid2008automated,10.1145/1961296.1950396} and
(later) greybox
\cite{10.1145/2976749.2978428,libFuzzer,christakis,10.1145/3368089.3417064}
fuzzing are approaches that use insights similar to those of symvalic
analysis, in an effort to achieve coverage of a program under test,
especially when the program makes use of very low level library code
that is externally modelled. These approaches work by ``fuzzing'' an
input, following the control flow of a program for concrete values,
yet also potentially using constraint solvers to modify the input to
follow alternative control flow. In the space of smart contracts,
where the program is fully modeled, and bugs manifest themselves in
several transactions, symvalic analysis can scale better and cover
more program behaviors, since it is fundamentally a static analysis,
overapproximating dynamic conditions and collapsing many paths.

Symvalic analysis is rather unconventional among analysis techniques,
mainly in the ways described earlier: it uses symbolic expressions
inside a full \emph{static} analysis (not dynamic-symbolic execution),
without delegating the solution of large expressions to an external
constraint solver. There have been other combinations of symbolic
expressions and static analyses, especially for custom analysis
clients---e.g., \citet{Dudina2017UsingSS} employ a symbolic static
analysis for buffer overflow detection. In contrast, symvalic analysis
is client-agnostic: symbolic expressions are used as regular values
for \emph{any} variable in the program text, without targeting
specific kinds of expressions or specific program
features. Furthermore, the mechanism of dependencies
(Section~\ref{sec:dependencies}) is key in the symvalic design, for
purposes of precision.

There are certainly many more points in the static analysis design
space and some can be compared for
reference. SPLlift~\cite{10.1145/2491956.2491976} shows a modular
analysis for software product lines. This is almost at the opposite
end of the spectrum from symvalic analysis: a very scalable analysis,
but much less precise. The SPLlift analysis is explicitly based on the
IFDS framework, which means that it summarizes at the procedure
boundary, thus losing precision to gain scalability. It is interesting
to consider whether symvalic analysis could apply to large-scale
software product lines. This direction would certainly require
significant work for a fruitful approach. For instance, symvalic
analysis would likely apply to each Java file separately with the
dependencies between different files modeled rather loosely (e.g.,
perhaps a single predicate to capture the configuration of the product
line). \citet{10.1145/2487568.2487569} describe an analysis that achieves
significant precision (flow- and context sensitivity) in a fairly general
analysis setting. This may be a promising candidate for future
combinations with symvalic analysis ideas, especially the use of
dependencies as a data-flow-value context.

\section{Conclusions}

We have argued that automated program analysis techniques can have a significant
impact on the security of smart contracts. This does not mean that program analysis
can counter most smart contract threats, nor that positive results come easy. To have actual
impact on security, we have needed to invent new program analysis techniques and combine them
with significant domain expertise and state/environment information.

Apart from the technical elements (symvalic analysis and corpus
analysis) of our approach, perhaps the most useful aspect is the
understanding of the changing nature of analysis precision and report
rates in a high-impact setting. Conventional research in program
analysis seems tuned to catching ``cheap'' bugs: numerous and
low-impact. Such an approach also makes for good metrics: a high
warning rate and high precision (i.e., true-positive rate) are
possible for cheap bugs, but entirely unrealistic over rare high-value
vulnerabilities. Yet coming up with ways to quantify the success of
``needle in a haystack'' searches will be an important element of
achieving real impact on blockchain (and not only) security.



\bibliography{testing-cc,references,references2,references3,extras,ptranalysis,proceedings,specs,tools}

\end{document}